\documentclass[aps,prd,twocolumn,nofootinbib,superscriptaddress,showkeys]{revtex4}
\usepackage{amssymb,amsmath,graphicx,enumerate}
\usepackage{subfigure}
\usepackage{dcolumn,booktabs}
\usepackage{longtable}
\usepackage[dvipdfm,colorlinks=true,citecolor=blue,pdfstartview=FitH]{hyperref}

\usepackage{color,framed} 
\definecolor{shadecolor}{rgb}{1,0,0} 

\def\bea{\begin{equation}}
\def\eea{\end{equation}}
\newcommand{\rt}{Regge trajectory}
\newcommand{\rts}{Regge trajectories}
\newcommand{\bfr}{{\bf r}}

\newcommand{\bfpa}{{|\bf p|}}
\newcommand{\gev}{{\rm GeV}}

\newcommand{\sse}{spinless Salpeter equation}

\begin{document}
\title{Regge trajectory relations for the universal description of the heavy-light systems: diquarks, mesons, baryons and tetraquarks}
\author{Jiao-Kai Chen}
\email{chenjk@sxnu.edu.cn, chenjkphy@outlook.com}
\affiliation{School of Physics and Information Engineering, Shanxi Normal University, Taiyuan 030031, China}

\begin{abstract}
Two newly proposed Regge trajectory relations are employed to analyze the heavy-light systems. One of the relations is $M=m_1+m_2+C'+\beta_x\sqrt{x+c_{0x}}$, $(x=l,\,n_r)$. Another reads $M=m_1+C'+\sqrt{\beta_x^2(x+c_{0x})+\frac{4}{3}\sqrt{{\pi}{\beta_x}}m^{3/2}_2(x+c_{0x})^{1/4}}$. $M$ is the bound state mass. $m_1$ and $m_2$ are the masses of the heavy constituent and the light constituent, respectively. $l$ is the orbital angular momentum and $n_r$ is the radial quantum number. $\beta_x$ and $c_{0x}$ are fitted. $m_1$, $m_2$ and $C'$ are input parameters. These two formulas consider both of the masses of heavy constituent and light constituent.
We find that the heavy-light diquarks, the heavy-light mesons, the heavy-light baryons and the heavy-light tetraquarks satisfy these two formulas.
When applying the first formula, the heavy-light systems satisfy the universal description irrespective of both of the masses of the light constituents and the heavy constituent. When using the second relation, the heavy-light systems satisfy the universal description irrespective of the mass of the heavy constituent.
The fitted slopes differ distinctively for the heavy-light mesons, baryons and tetraquarks, respectively. When employing the first relation, the average values of $c_{fn_r}$ ($c_{fl}$) are $1.026$, $0.794$ and $0.553$ ($1.026$, $0.749$ and $0.579$) for the heavy-light mesons, the heavy-light baryons and the heavy-light tetraquarks, respectively. Upon application of the second relation, the mean values of $c_{fn_r}$ ($c_{fl}$) are $1.108$, $0.896$ and $0.647$ ($1.114$, $0.855$ and $0.676$) for the heavy-light mesons, the heavy-light baryons and the heavy-light tetraquarks, respectively. Moreover, the fitted results show that the Regge trajectories for the heavy-light systems are concave downwards in the $(M^2,\,n_r)$ and $(M^2,\,l)$ planes.
\end{abstract}


\maketitle


\section{Introduction}
There is a vast amount of experimental data on the spectra of different types of hadrons \cite{ParticleDataGroup:2022pth}. One of the widely used approaches in studying the hadron spectra is the {\rt} \cite{Regge:1959mz,Collins:1971ff,Collins:1977jy,Gross:2022hyw,Inopin:2001ub,
Inopin:1999nf,Chew:1961ev,Chew:1962eu,Guo:2008he,Feng:2022hpj,MartinContreras:2023oqs,
Brisudova:1999ut,Sergeenko:1994ck,
Veseli:1996gy,Afonin:2014nya,Burns:2010qq,MartinContreras:2020cyg,
Baldicchi:1998gt,Tang:2000tb,Badalian:2019lyz,Chen:2018nnr}. The study of {\rts} of hadron spectra is beneficial for gaining insights into the strong interactions from various perspectives \cite{Chen:2016spr,Klempt:2012fy,
Brodsky:2018vyy,Nielsen:2018uyn,Chen:2017fcs,Sonnenschein:2018fph,
MartinContreras:2020cyg,Sharov:2013tga,Forkel:2007cm,A:2023bxv}. Clearly, an unified dynamic mechanism will lead to the universal description of {\rts} for hadrons \cite{Brodsky:2018vyy,Nielsen:2018uyn,Sonnenschein:2018fph}.

The meson {\rts} exhibit structures that vary in different energy regions \cite{Chen:2021kfw,Chen:2022flh}. This variability is expected to hold true for other hadrons with a similar dynamic mechanism, such as baryons and tetraquarks in the diquark picture.
For the light-light systems, it is well-known that they can be well described by the renowned linear {\rts} \cite{Chew:1961ev,Chew:1962eu}.
In case of the heavy-heavy systems, an universal description is presented in Refs. \cite{Chen:2023djq,Feng:2023txx}. Regarding the heavy-light systems, the authors provide an universal description of the orbitally excited heavy-light mesons and baryons in Ref. \cite{Chen:2017fcs}.
In Ref. \cite{Chen:2023cws}, we present the heavy-light diquark {\rts}. The proposed {\rt} relations can universally describe the heavy-light mesons and the heavy-light diquarks.
In this study, we apply the {\rt} relations \footnote{The {\rts} of hadrons are commonly plotted in the $(M^2,\,x)$ plane or in the $(x,\,M^2)$ plane, where $x=l,\,n_r$. For simplicity, the figures plotted in the $(M,\,x)$ plane and in the $((M-m_R)^2,\,x)$ plane are also called the {\rts}. In this work, we concentrate on the $\lambda$-mode of baryons and tetraquarks and the $\rho$-mode excitations of diquarks are not considered.} newly proposed in Ref. \cite{Chen:2023cws} to the heavy-light baryons composed of a heavy quark and a light diquark or composed of a light quark and a heavy diquark, as well as to the heavy-light tetraquarks consisting of one heavy (light) diquark and one light (heavy) antidiquark.

This paper is organized as follows: In Sec. \ref{sec:rt}, the {\rt} relations are obtained from the {\sse} (SSE). In Sec. \ref{sec:fit}, the universal description of the heavy-light diquarks, mesons, baryons and tetraquarks is investigated. Discussions on concavity of the {\rts} are given in Sec. \ref{sec:partners} and the conclusions are in Sec. \ref{sec:concl}.

\section{{\rt} relations for the heavy-light systems}\label{sec:rt}

\subsection{SSE}
The {\sse} (SSE) \cite{Godfrey:1985xj,Ferretti:2019zyh,Bedolla:2019zwg,Durand:1981my,Durand:1983bg,Lichtenberg:1982jp,Jacobs:1986gv} reads as
\begin{eqnarray}\label{qsse}
M\Psi_{d,m,b,t}({\bfr})=M_0\Psi_{d,m,b,t}({\bfr})+V_{d,m,b,t}\Psi_{d,m,b,t}({\bfr}),
\end{eqnarray}
where $M_0=\omega_1+\omega_2$. $M$ is the bound state mass (diquark, meson, baryon and tetraquark). $\Psi_{d,m,b,t}({\bfr})$ are the diquark wave function, the meson wave function, the baryon wave function, and the tetraquark wave function, respectively. $V_{d,m,b,t}$ are the diquark potential, the meson potential, the baryon potential, and the tetraquark potential, respectively. $\omega_1$ is the relativistic energy of constituent $1$, and $\omega_2$ is of constituent $2$,
\bea\label{omega}
\omega_i=\sqrt{m_i^2+{\bf p}^2}=\sqrt{m_i^2-\Delta}\;\; (i=1,2).
\eea
$m_1$ and $m_2$ are the effective masses of heavy constituent $1$ and light constituent $2$, respectively.

In the present work, the Cornell-like potential is considered \cite{Eichten:1974af,Ferretti:2019zyh,Bedolla:2019zwg,Ferretti:2011zz},
\bea\label{potv}
V_{d,m,b,t}=-\frac{3}{4}\left[V_c+{\sigma}r+C\right]
({\bf{F}}_i\cdot{\bf{F}}_j)_{d,m,b,t},
\eea
where $V_c\,{\propto}\,1/r$ is the color Coulomb potential or a Coulomb-like interaction for different hadrons \cite{Ferretti:2019zyh,Ferretti:2011zz}. The second term is the linear confining potential and $\sigma$ is the string tension. $C$ is a fundamental parameter \cite{Gromes:1981cb,Lucha:1991vn}. ${\bf{F}}_i\cdot{\bf{F}}_j$ is the color-Casmir,
\bea
\langle({\bf{F}}_i\cdot{\bf{F}}_j)_{d}\rangle=-\frac{2}{3},\quad
\langle({\bf{F}}_i\cdot{\bf{F}}_j)_{m,b,t}\rangle=-\frac{4}{3}.
\eea

From Eqs. (\ref{qsse}) and (\ref{potv}), we see that the heavy-light diquarks, the heavy-light mesons, the heavy-light baryons, and the heavy-light tetraquarks are described in an unified approach \cite{Ferretti:2019zyh,Bedolla:2019zwg}. Therefore, it is expected that these heavy-light systems can be described universally by the {\rt} approach.

\subsection{{\rt} relations}
The mass of the light constituent is assumed to approach zero, $m_2\to0$ in Refs. \cite{Veseli:1996gy,Chen:2017fcs} or is considered by correction term in Refs. \cite{Selem:2006nd,Chen:2014nyo,Afonin:2014nya,Sonnenschein:2018fph,Jakhad:2023mni}. In the limit $m_1\to\infty$ and $m_2\to0$, Eq. (\ref{qsse}) is reduced to be
\begin{eqnarray}\label{qssenr}
M\Psi_{d,m,b,t}({\bfr})=\left[m_1+{\bfpa}+V_{d,m,b,t}\right]\Psi_{d,m,b,t}({\bfr}).
\end{eqnarray}
By employing the Bohr-Sommerfeld quantization approach \cite{brau:04bs,brsom}, we have from Eq. (\ref{qssenr})
\bea\label{rtfsm}
M{\sim}2\sqrt{\sigma'}\sqrt{l},\;M{\sim}\sqrt{2{\pi}\sigma'}\sqrt{n_r},
\eea
where
\bea
\sigma'=\left\{\begin{array}{cc}
\sigma/2, & \text{diquarks}, \\
\sigma, & \text{mesons, baryons, tetraquarks}.
\end{array}\right.
\eea
Using Eq. (\ref{rtfsm}), the parameterized formula can be written as \cite{Chen:2023cws}
\bea\label{rtmeson}
M=m_R+\beta_x\sqrt{x+c_{0x}},\,(x=l,\,n_r).
\eea
The parameter in Eq. (\ref{rtmeson}) reads as \cite{Chen:2021kfw}
\bea\label{massform}
\beta_x=c_{fx}c_xc_{c}.
\eea
The constants $c_{x}$ [$c_{n_r}$ and $c_l$] and $c_{c}$ are
\bea\label{cxcons}
c_{c}=\sqrt{\sigma'},\quad c_l=2,\quad c_{n_r}=\sqrt{2\pi}.
\eea
Both $c_{fl}$ and $c_{fn_r}$ are theoretically equal to one and are fitted in practice. For the heavy-light mesons, the common choice of $m_R$ is \cite{Selem:2006nd,Chen:2021kfw,Jakhad:2023mni,Chen:2014nyo,
Chen:2017fcs,Veseli:1996gy,Jia:2018vwl}
\bea\label{mrm1}
m_R=m_1.
\eea
$m_R$, $c_x$ and $\sigma$ are universal for the heavy-light systems. $c_{fx}$ and $c_{0x}$, which vary with different {\rts}, are determined by fitting the given {\rt}.

The usual {\rt} Eq. (\ref{rtmeson}) with (\ref{mrm1}), which is obtained in the limit $m_1\to\infty$ and $m_2\to0$, cannot give agreeable results of the heavy-light diquarks.
Corresponding to different ways to include the light constituent's mass, two modified formulas are proposed in Ref. \cite{Chen:2023cws}, which can describe universally both the heavy-light mesons and the heavy-light diquarks. One is
Eq. (\ref{rtmeson}) with
\bea\label{rtft}
m_R=m_1+m_2+C',
\eea
where
\bea
C'=\left\{\begin{array}{cc}
C/2, & \text{diquarks}, \\
C, & \text{mesons, baryons, tetraquarks}.
\end{array}\right.
\eea
Another reads
\bea\label{mrtf}
M=m_R+\sqrt{\beta_x^2(x+c_{0x})+\kappa_{x}m^{3/2}_2(x+c_{0x})^{1/4}}
\eea
if $m_2{\ll}M$, where
\bea\label{mrfp}
m_R=m_1+C',\quad \kappa_x=\frac{4}{3}\sqrt{{\pi}\beta_x},
\eea
where $\beta_x$ is in (\ref{massform}).
(\ref{rtmeson}) with (\ref{rtft}) is an extension of $M=m_1+m_2+\sqrt{a(n_r+{\alpha}l+b)}$ \cite{Afonin:2014nya}, while (\ref{mrtf}) with (\ref{mrfp}) is based on the results in \cite{Selem:2006nd,Sonnenschein:2018fph}.
As $m_2=0$, these two modified formulas, formulas (\ref{rtmeson}) with (\ref{rtft}) and (\ref{mrtf}) with (\ref{mrfp}), become identical. As $m_2=0$ and $C$ is neglected, these two modified formulas reduce to the usual {\rt} formula for the heavy-light mesons, i.e., (\ref{rtmeson}) with (\ref{mrm1}).
In this work, we apply these two modified formulas to the heavy-light systems: diquarks, mesons, baryons, tetraquarks.

\section{Universal description of the heavy-light systems}\label{sec:fit}

In this section we present the universal description of different types of heavy-light systems by employing (\ref{rtmeson}) with (\ref{rtft}) and (\ref{mrtf}) with (\ref{mrfp}). The {\rts} for the heavy-light systems are fitted individually.

\subsection{Parameters}
The parameters are \cite{Ebert:2007rn,Ebert:2011kk,Ebert:2009ua}
\begin{align}\label{paramet}
&m_u=m_d=0.33\, {\gev},\quad m_s=0.50\, {\gev},\nonumber\\
&m_c=1.55\, {\gev},\quad m_b=4.88\, {\gev},\nonumber\\
& \sigma=0.18\,{\gev^2},\quad C=-0.3\,{\gev},  \nonumber\\
& m_{[ud]}=0.710\,{\gev},\quad m_{\{ud\}}=0.909\,{\gev},\nonumber\\
& m_{[us]}=0.948\,{\gev},\quad m_{\{us\}}=1.069\,{\gev},\nonumber\\
& m_{\{cc\}}=3.226\,{\gev},\quad  m_{\{bb\}}=9.778\,{\gev},\nonumber\\
& m_{\{cb\}}=6.526\,{\gev},\quad  m_{\{ss\}}=1.203\,{\gev},
\end{align}
where $\{\;\}$ denotes the axial-vector diquark and $[\;]$ the scalar diquark.
$c_{0x}$ and $c_{fx}$ vary with different {\rts}.

The values in Eq. (\ref{paramet}) are used to calculate the masses of tetraquarks \cite{Ebert:2007rn}, baryons \cite{Ebert:2011kk} and mesons \cite{Ebert:2009ua}.
In Ref. \cite{Chen:2017fcs}, the values are used to give an universal description of the heavy-light mesons and baryons.
The values are also used to discuss diquarks \cite{Feng:2023txx,Chen:2023cws}.

\subsection{Heavy-light diquarks}

Diquarks have been discussed by various approaches, for example, the QCD sum rules \cite{Dosch:1988hu,Zhang:2006xp,Jamin:1989hh,deOliveira:2023hma,Kleiv:2013dta,
Wang:2011ab,Esau:2019hqw,Wang:2010sh}.
In Ref. \cite{Chen:2023cws}, we show that the Regge trajectories for
the heavy-light diquarks can be well described by (\ref{rtmeson}) with (\ref{rtft}) and (\ref{mrtf}) with (\ref{mrfp}). The spectra of the heavy-light diquarks
$(cu)$, $(cs)$, $(bu)$ and $(bs)$ obtained by using the Regge trajectory
approach agree with other theoretical predictions. See \cite{Chen:2023cws} for more details.

In case of Eq. (\ref{rtmeson}) with (\ref{rtft}), the heavy-light diquarks satisfy the universal description irrespective of both mass of the light constituents and mass of the heavy constituent. In case of Eq. (\ref{mrtf}) with (\ref{mrfp}), the heavy-light diquarks satisfy the universal description irrespective of mass of the heavy constituents.

\subsection{Heavy-light mesons}

\begin{table*}[!phtb]
\caption{The experimental values \cite{ParticleDataGroup:2022pth} and the theoretical values (EFG) \cite{Ebert:2009ua} for the charmed and bottom mesons. The values are in {\gev}. }  \label{tab:mm}
\centering
\begin{tabular*}{0.9\textwidth}{@{\extracolsep{\fill}}ccccccc@{}}
\hline\hline
State         &    Meson         & PDG      & EFG            & Meson  & PDG  & EFG  \\
\hline
$1^3S_1$       &$D^{\ast}(2010)^{\pm}$  & $2.01026$ & 2.010  & $D^{\ast\pm}_s$ &2.1122 & 2.111     \\
$2^3S_1$ &$D^{\ast}_1(2600)^0$          & 2.627     & 2.632  & $D^{\ast}_{s1}(2700)^{\pm}$ &2.714 & 2.731 \\
$3^3S_1$       &                        &           & 3.096  &  &  &3.242    \\
$4^3S_1$       &                        &           & 3.482  &  &  &3.669  \\
$5^3S_1$       &                        &           & 3.822  &  &  &4.048    \\
$1^3P_2$       & $D^\ast_2(2460)$       & 2.4611   & 2.460   &$D^{\ast}_{s2}(2573)$  &2.5691  &2.571  \\
$1^3D_3$       & $D^\ast_3(2750)$       & 2.7631   & 2.863   &$D^{\ast}_{s3}(2860)^{\pm}$  &2.860  &2.971  \\
$1^3F_4$       &                        &          & 3.187   &  &  &3.300 \\
$1^3G_5$       &                        &          & 3.473   &  &  &3.595    \\
\hline
\end{tabular*}
\begin{tabular*}{0.9\textwidth}{@{\extracolsep{\fill}}ccccccc@{}}
\hline
State         &    Meson      & PDG    & EFG      &    Meson      & PDG    & EFG\\
\hline
$1^3S_1$       &$B^{\ast}$    &$5.32471$  & 5.326 &  $B^{\ast}_s$ &5.4154  &5.414    \\
$2^3S_1$       &              &           & 5.906 &  &  & 5.992    \\
$3^3S_1$       &              &           & 6.387 &  &  & 6.475      \\
$4^3S_1$       &               &          & 6.786 &  &  & 6.879   \\
$5^3S_1$       &               &          & 7.133 &  &  & 7.235     \\
$1^3P_2$       & $B^\ast_2(5747)$ &5.7372 & 5.741 &$B^{\ast}_{s2}(5840)^0$  &5.83986  &5.842   \\
$1^3D_3$       &                 &        & 6.091 &  &  &6.191   \\
$1^3F_4$       &                 &        & 6.380  &  &  &6.475  \\
$1^3G_5$       &                 &         & 6.634 &  &  &6.726    \\
\hline\hline
\end{tabular*}
\end{table*}

\begin{table*}[!phtb]
\caption{The fitted values of parameters $c_{fn_r}$, $c_{fl}$, $c_{0n_r}$, and $c_{0l}$ by employing formulas (\ref{rtmeson}) with (\ref{rtft}) (Fit1) and (\ref{mrtf}) with (\ref{mrfp}) (Fit2). The used data are in Tables \ref{tab:mm}, \ref{tab:bm}, \ref{tab:bmt} and \ref{tab:tm}.}  \label{tab:fitparameters}
\centering
\begin{tabular*}{0.9\textwidth}{@{\extracolsep{\fill}}ccccc@{}}
\hline\hline
           &   \multicolumn{2}{c}{Radial Traj.} &   \multicolumn{2}{c}{Orbital Traj.}    \\
                   & Fit1 &  Fit2  & Fit1   &  Fit2   \\
     &$(c_{fn_r},\,c_{0n_r})$  & $(c_{fn_r},\,c_{0n_r})$ & $(c_{fl},\,c_{0l})$ & $(c_{fl},\,c_{0l})$  \\
\hline
$D^{\ast}(2010)^{\pm}$  &(1.038,\,0.008)  &(1.103,\,0.17) &(1.093,\,0.039) &(1.166,\,0.23)     \\
$D_s^{\ast\pm}$      &(1.053,\,0.0)   &(1.152,\,0.105) &(1.065,\,0.0) &(1.168,\,0.165)     \\
$B^{\ast}$             &(1.025,\,0.0) &(1.092,\,0.15) &(0.991,\,0.091) &(1.067,\,0.26)     \\
$B_s^{\ast}$           &(0.986,\,0.0) &(1.085,\,0.1) & (0.955,\,0.005)& (1.054,\,0.175)     \\
$\Lambda_c^{+}$        &(0.827,\,0.0)&(0.928,\,0.15) &(0.773,\,0.076) &(0.875,\,0.275)     \\
$\Lambda_b^0$          &(0.812,\,0.0)&(0.912,\,0.155)&(0.762,\,0.093) &(0.864,\,0.285)     \\
$\Sigma_c$             &(0.789,\,0.0)&(0.888,\,0.21)&(0.693,\,0.235) &(0.792,\,0.44)     \\
$\Sigma_b$             &(0.757,\,0.0)&(0.854,\,0.205)&(0.643,\,0.232) &(0.732,\,0.475)     \\
$\Xi_c$                &(0.763,\,0.0)&(0.863,\,0.205)&(0.692,\,0.161)& (0.788,\,0.415)     \\
$\Xi_b$                &(0.730,\,0.0)&(0.827,\,0.210)&(0.615,\,0.280) &(0.710,\,0.495)     \\
$\Omega_c$             &(0.756,\,0.0)&(0.856,\,0.205)&(0.692,\,0.135) &(0.786,\,0.425)     \\
$\Omega_b$             &(0.718,\,0.0)&(0.814,\,0.225)&(0.624,\,0.149) &(0.705,\,0.50)     \\
$\Xi_{cc}$             &             &               &(0.902,\,0.379) &(1.020,\,0.410)     \\
$\Xi_{bb}$             &(0.903,\,0.200) &(1.015,\,0.22)&(0.869,\,0.338) &(0.984,\,0.37)     \\
$\Omega_{cc}$          &                &              &(0.890,\,0.349) &(1.032,\,0.37)     \\
$\Omega_{bb}$          &(0.872,\,0.196)&(1.004,\,0.215)&(0.836,\,0.336) &(0.971,\,0.36)     \\
$\{cc\}[\bar{u}\bar{d}]$ &(0.611,\,0.251)&(0.718,\,0.285)&(0.618,\,0.384) &(0.726,\,0.44)     \\
$\{bb\}[\bar{u}\bar{d}]$ &(0.517,\,0.300)&(0.606,\,0.355)&(0.598,\,0.352) &(0.701,\,0.415)     \\
$\{cc\}[\bar{n}\bar{s}]$ &(0.579,\,0.188)&(0.677,\,0.27)&(0.548,\,0.329) &(0.637,\,0.48)     \\
$\{bb\}[\bar{n}\bar{s}]$ &(0.484,\,0.214)&(0.564,\,0.335)&(0.521,\,0.290) &(0.605,\,0.46)     \\
$\{cb\}[\bar{u}\bar{d}]$  &(0.563,\,0.268)&(0.660,\,0.315)&(0.618,\,0.349) &(0.724,\,0.41)     \\
$\{cb\}[\bar{n}\bar{s}]$  &(0.528,\,0.196)&(0.615,\,0.300)&(0.553,\,0.281) &(0.643,\,0.430)     \\
$\{cc\}\{\bar{n}\bar{n}\}$&(0.614,\,0.233)&(0.718,\,0.300)&(0.594,\,0.393) &(0.694,\,0.505)     \\
$\{bb\}\{\bar{n}\bar{n}\}$&(0.521,\,0.289)&(0.608,\,0.385)&(0.564,\,0.388) &(0.658,\,0.515)     \\
$\{cb\}\{\bar{n}\bar{n}\}$&(0.564,\,0.260)&(0.659,\,0.34)&(0.599,\,0.362) &(0.699,\,0.475)     \\
\hline
\hline
\end{tabular*}
\end{table*}

\begin{figure*}[!phtb]
\centering
\subfigure[]{\label{subfigure:fiterr}\includegraphics[scale=0.45]{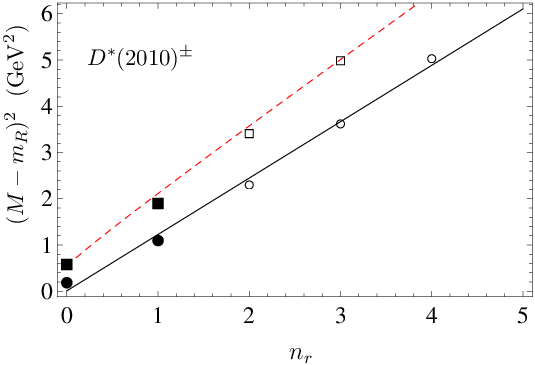}}
\subfigure[]{\label{subfigure:fiterr}\includegraphics[scale=0.45]{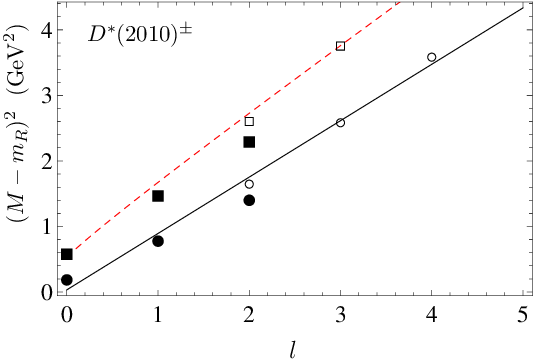}}
\subfigure[]{\label{subfigure:fiterr}\includegraphics[scale=0.45]{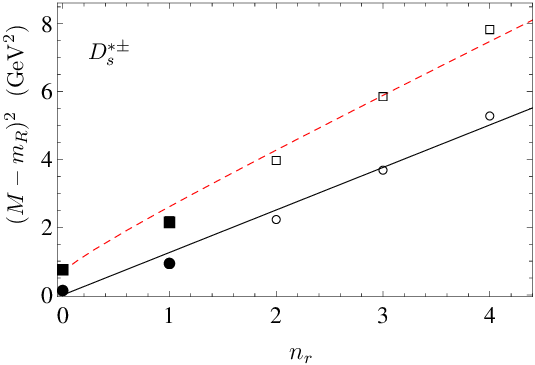}}
\subfigure[]{\label{subfigure:fiterr}\includegraphics[scale=0.45]{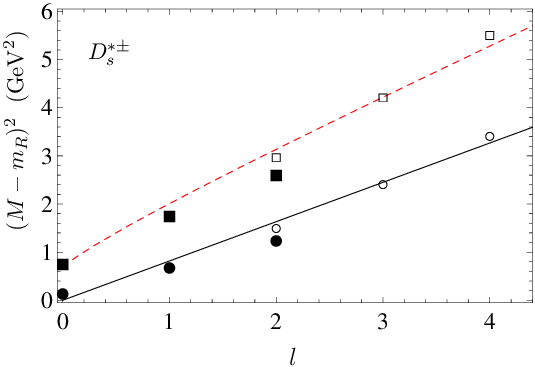}}
\subfigure[]{\label{subfigure:fiterr}\includegraphics[scale=0.45]{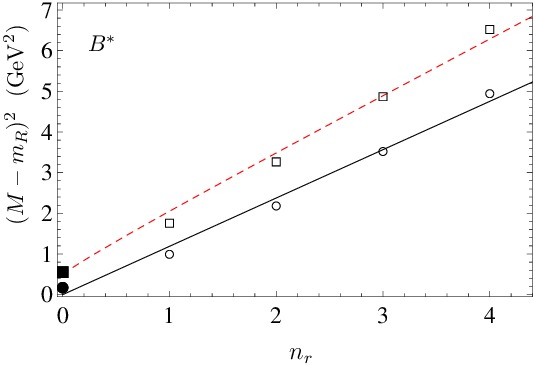}}
\subfigure[]{\label{subfigure:fiterr}\includegraphics[scale=0.45]{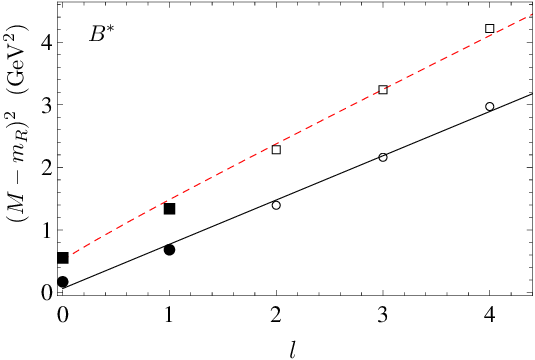}}
\subfigure[]{\label{subfigure:fiterr}\includegraphics[scale=0.45]{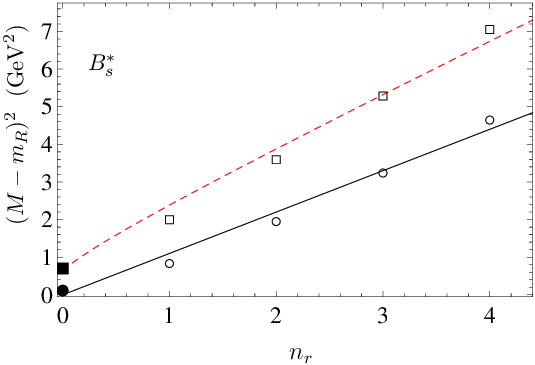}}
\subfigure[]{\label{subfigure:fiterr}\includegraphics[scale=0.45]{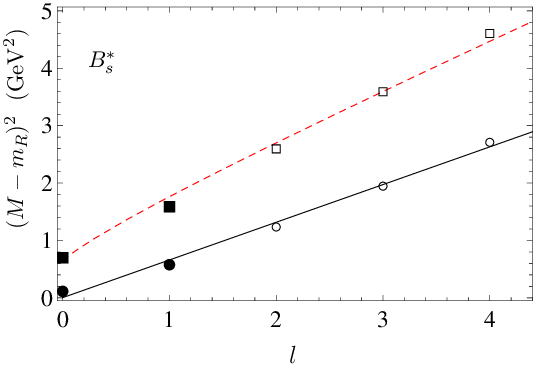}}
\caption{The radial and orbital {\rts} for the heavy-light mesons by employing formulas (\ref{rtmeson}) with (\ref{rtft}) (the black lines) and (\ref{mrtf}) with (\ref{mrfp}) (the red dashed lines). The PDG data (the dots and the filled squares) and the theoretical data (the circles and the empty squares) are listed in Table \ref{tab:mm}. $m_R$ for the black lines and for the red dashed lines are different, see (\ref{rtft}) and (\ref{mrfp}).}\label{fig:mr}
\end{figure*}

\begin{figure*}[!phtb]
\centering
\subfigure[]{\label{subfigure:fiterr}\includegraphics[scale=0.47]{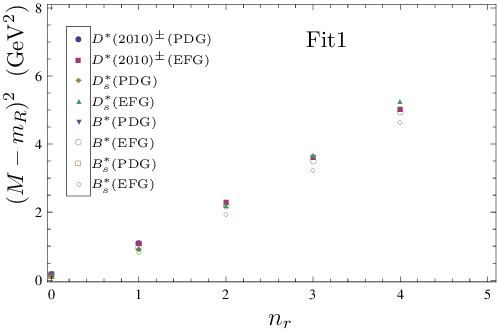}}
\subfigure[]{\label{subfigure:fiterr}\includegraphics[scale=0.47]{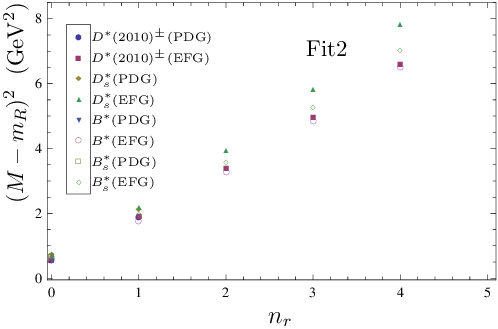}}
\subfigure[]{\label{subfigure:fiterr}\includegraphics[scale=0.47]{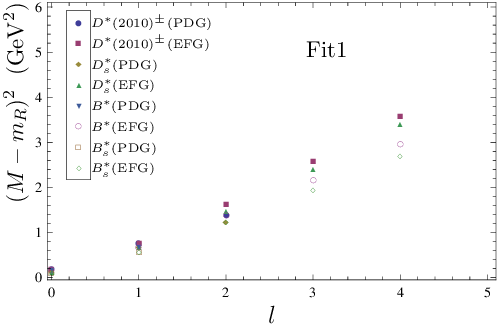}}
\subfigure[]{\label{subfigure:fiterr}\includegraphics[scale=0.47]{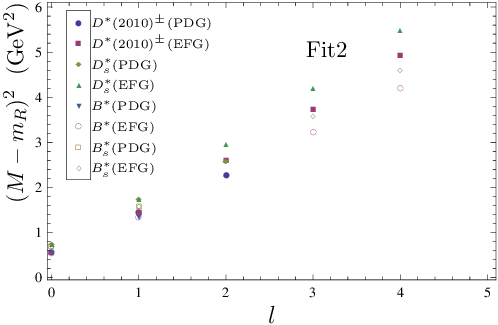}}
\caption{The heavy-light mesons are ploted in the $((M-m_1-m_2-C)^2,\,x)$ plane (Fit1) and in the $((M-m_1-C)^2,\,x)$ plane (Fit2), respectively. $x=l,\,n_r$. $m_1$ is the mass of the heavy constituent and $m_2$ the mass of the light constituent. $C$ is in (\ref{potv}). The used data are listed in Table \ref{tab:mm}. }\label{fig:mc}
\end{figure*}

By using Eqs. (\ref{rtmeson}) with (\ref{rtft}) and (\ref{mrtf}) with (\ref{mrfp}) and data in Table \ref{tab:mm}, the radial and orbital {\rts} for the heavy-light mesons $D^{\ast}$, $D_s^{\ast\pm}$, $B^{\ast}$ and $B^{\ast}_s$ are obtained, see Fig. \ref{fig:mr}. The fitted values are listed in Table \ref{tab:fitparameters}.
In Fig. \ref{fig:mr}, the {\rts} (the red dashed lines) obtained by employing (\ref{mrtf}) with (\ref{mrfp}) lie above the {\rts} (the black lines) obtained by applying (\ref{rtmeson}) with (\ref{rtft}). It is because $m_R$ for these two {\rts} are different, see Eqs. (\ref{rtft}) and (\ref{mrfp}). The figures show that the heavy-light mesons satisfy both of these two {\rt} relations.

From Fig. \ref{fig:mc} and the fitted values in Table \ref{tab:fitparameters}, we can see that both of the radial and orbital {\rts} in the $((M-m_1-m_2-C)^2,\,x)$ planes almost overlap with each other irrespective of both the light quark flavor and heavy quark flavor. It shows an universal description of these heavy-light mesons.
For the heavy-light mesons, the {\rts} in the $((M-m_1-C)^2,\,x)$ plane are universal only irrespective of heavy quark flavor.

As employing Eq. (\ref{rtmeson}) with (\ref{rtft}), the average values of $c_{fn_r}$ and $c_{fl}$ are $1.026$ and $1.026$ for the heavy-light mesons, respectively. When applying Eq. (\ref{mrtf}) with (\ref{mrfp}), the average values of $c_{fn_r}$ and $c_{fl}$ are $1.108$ and $1.114$, respectively.

From Table \ref{tab:fitparameters}, we can see that the fitted values of $c_{fn_r}$ and $c_{fl}$ decrease in the majority of cases as the masses of the constituents increase when employing (\ref{rtmeson}) with (\ref{rtft}). However, the fitted values of $c_{fn_r}$ and $c_{fl}$ obtained by using (\ref{mrtf}) with (\ref{mrfp}) frequently decrease with the increase of the masses of the heavy constituent and rise with the increase of the masses of the light constituent. This trend aligns with the results in \cite{Chen:2017fcs}.

\subsection{Heavy-light baryons}

\begin{table*}[!phtb]
\caption{The experimental and theoretical data for the radially excited states $\frac{1}{2}^+$ of baryons $\Lambda_c^+$ and $\Lambda_b^0$. Q denotes the heavy quark. The values are in {\gev}. }  \label{tab:bm}
\centering
\begin{tabular*}{0.95\textwidth}{@{\extracolsep{\fill}}clllllc@{}}
\hline\hline
Q[ud] State         &    Baryon  & PDG \cite{ParticleDataGroup:2022pth}      & EFG \cite{Ebert:2011kk}
    & Baryon  & PDG \cite{ParticleDataGroup:2022pth}      & EFG \cite{Ebert:2011kk}     \\
\hline
$1S$       & $\Lambda_c^{+}$   & 2.28646  & 2.286
    & $\Lambda_b^0$  & 5.61960  & 5.620   \\
$2S$       &   &   & 2.769
    &               &          & 6.089  \\
$3S$       &   &   & 3.130
    &               &          & 6.455    \\
$4S$       &   &   & 3.437
    &               &          & 6.756    \\
$5S$       &   &   & 3.715
    &               &          & 7.015     \\
$6S$       &   &   & 3.973
    &               &          & 7.256   \\
$1P$       &$\Lambda_c(2625)^{+}$ &2.62811 &2.627
    &                &          & 5.942    \\
$1D$       &$\Lambda_c(2880)^{+}$ & 2.88163  &2.880
    &                &          & 6.196  \\
$1F$       &                      &        & 3.078
   &                &          & 6.411   \\
$1G$       &                      &        & 3.284
    &                &          & 6.599     \\
$1H$       &                      &        & 3.460
    &                &          & 6.766      \\
\hline
\end{tabular*}
\end{table*}

\begin{figure*}[!phtb]
\centering
\subfigure[]{\label{subfigure:fiterr}\includegraphics[scale=0.45]{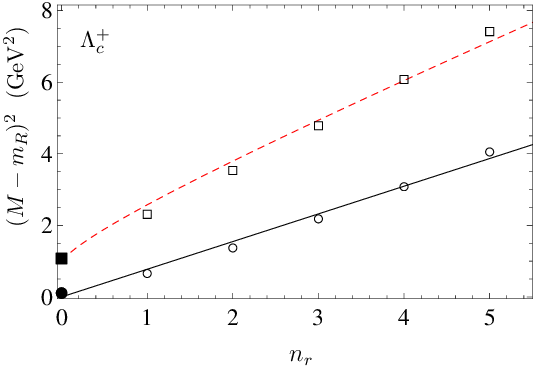}}
\subfigure[]{\label{subfigure:fiterr}\includegraphics[scale=0.45]{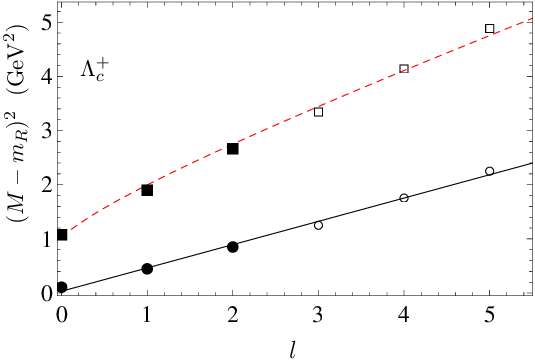}}
\subfigure[]{\label{subfigure:fiterr}\includegraphics[scale=0.45]{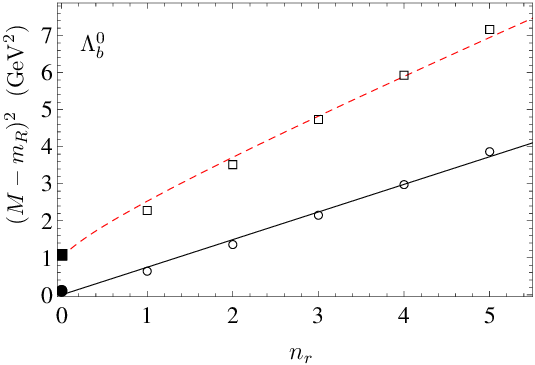}}
\subfigure[]{\label{subfigure:fiterr}\includegraphics[scale=0.45]{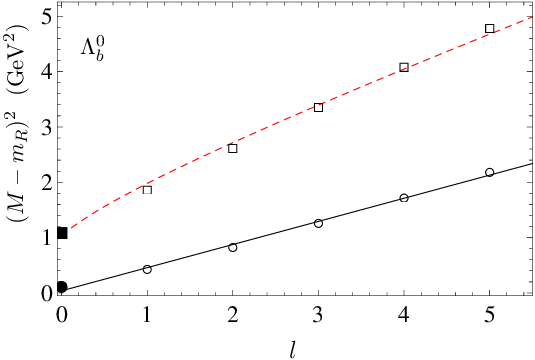}}
\subfigure[]{\label{subfigure:fiterr}\includegraphics[scale=0.45]{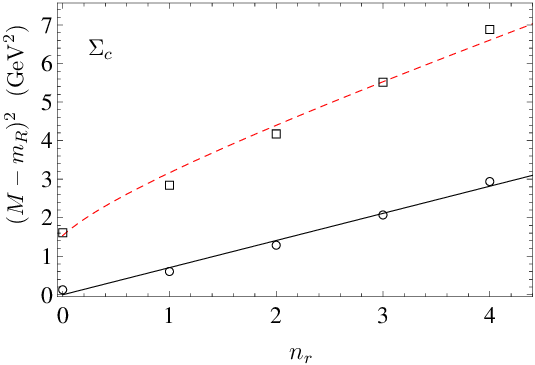}}
\subfigure[]{\label{subfigure:fiterr}\includegraphics[scale=0.45]{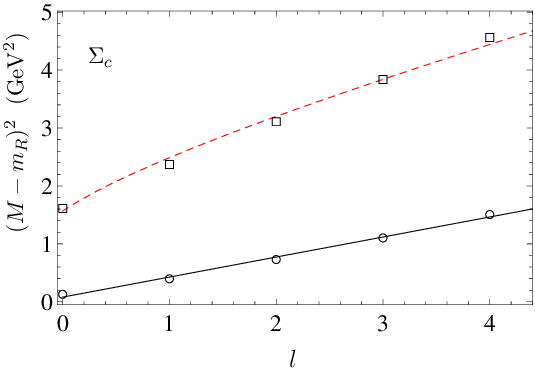}}
\subfigure[]{\label{subfigure:fiterr}\includegraphics[scale=0.45]{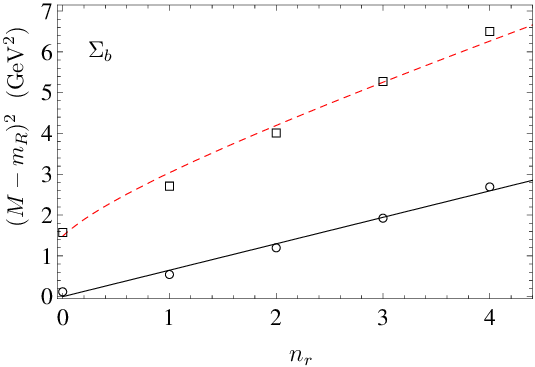}}
\subfigure[]{\label{subfigure:fiterr}\includegraphics[scale=0.45]{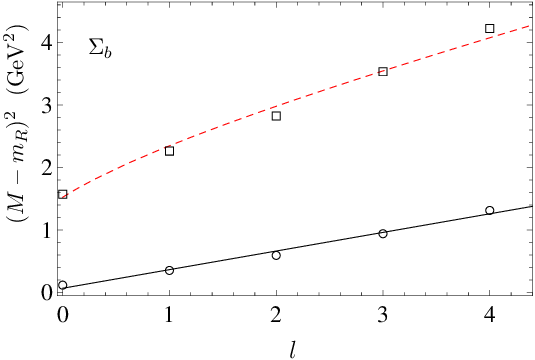}}
\subfigure[]{\label{subfigure:fiterr}\includegraphics[scale=0.45]{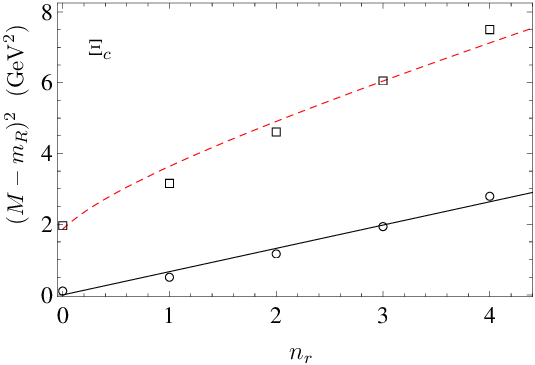}}
\subfigure[]{\label{subfigure:fiterr}\includegraphics[scale=0.45]{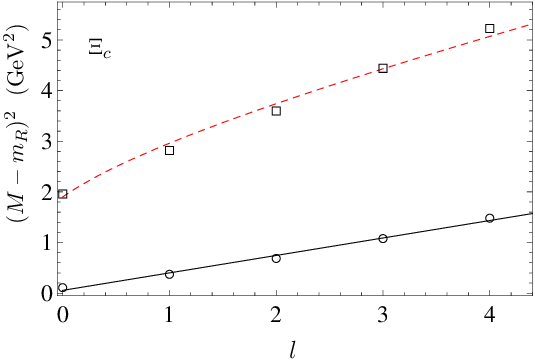}}
\subfigure[]{\label{subfigure:fiterr}\includegraphics[scale=0.45]{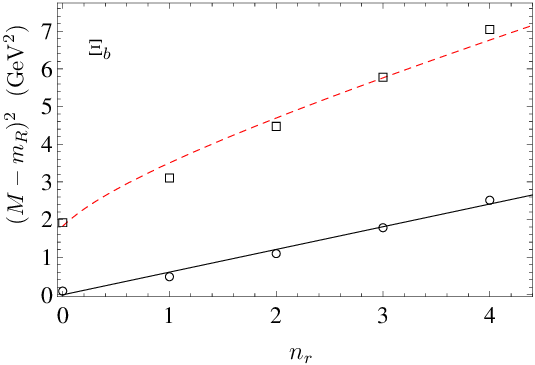}}
\subfigure[]{\label{subfigure:fiterr}\includegraphics[scale=0.45]{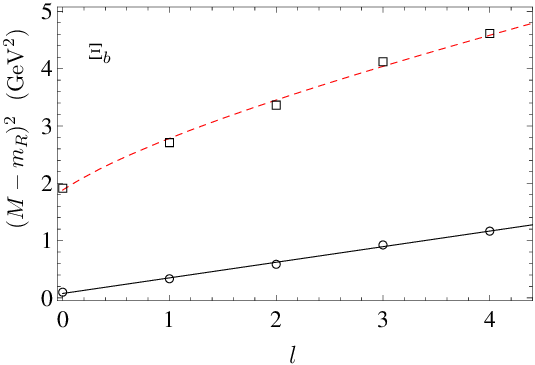}}
\subfigure[]{\label{subfigure:bocr}\includegraphics[scale=0.45]{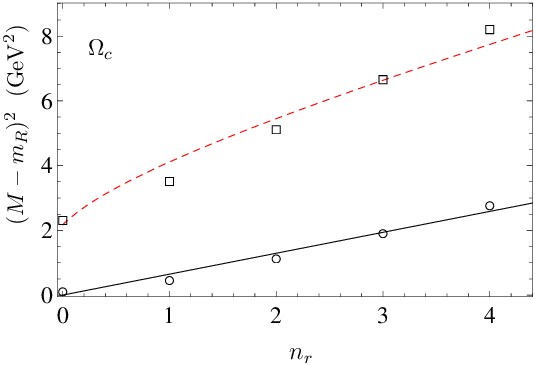}}
\subfigure[]{\label{subfigure:fiterr}\includegraphics[scale=0.45]{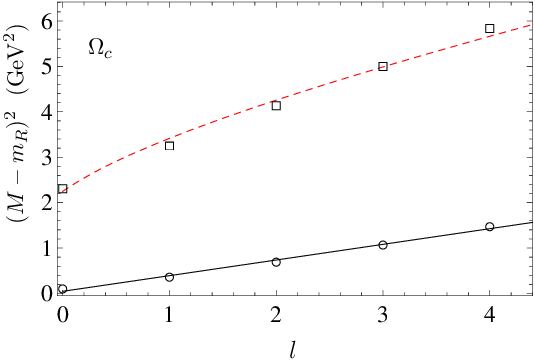}}
\subfigure[]{\label{subfigure:fiterr}\includegraphics[scale=0.45]{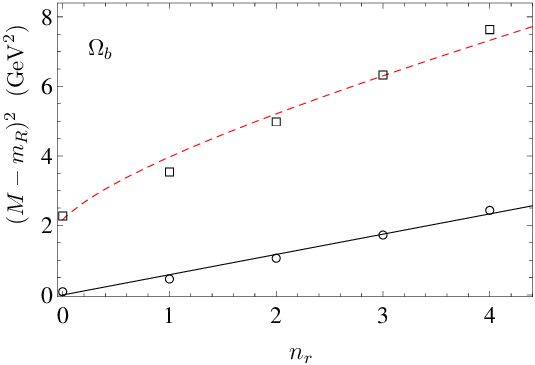}}
\subfigure[]{\label{subfigure:fiterr}\includegraphics[scale=0.45]{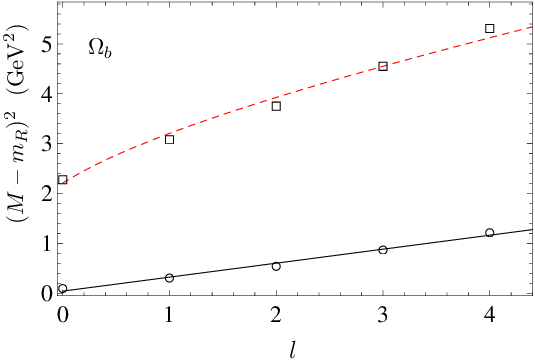}}
\subfigure[]{\label{subfigure:fiterr}\includegraphics[scale=0.45]{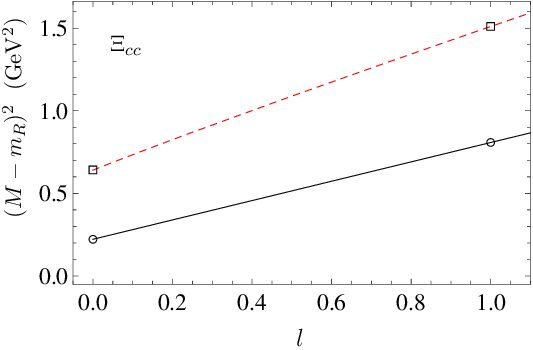}}
\subfigure[]{\label{subfigure:fiterr}\includegraphics[scale=0.45]{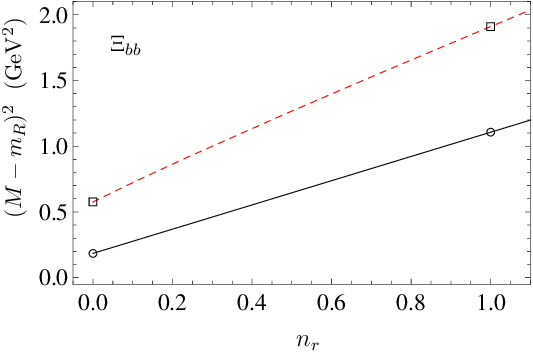}}
\subfigure[]{\label{subfigure:fiterr}\includegraphics[scale=0.45]{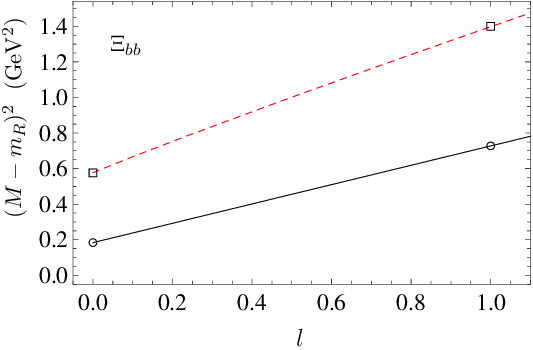}}
\subfigure[]{\label{subfigure:fiterr}\includegraphics[scale=0.45]{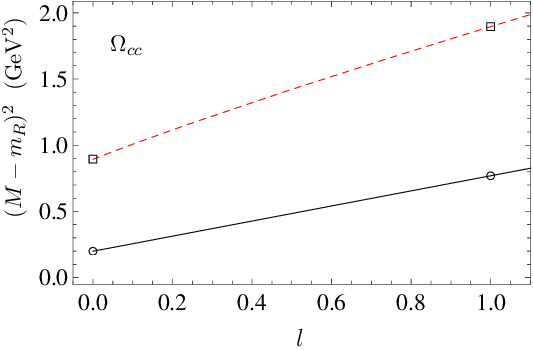}}
\subfigure[]{\label{subfigure:fiterr}\includegraphics[scale=0.45]{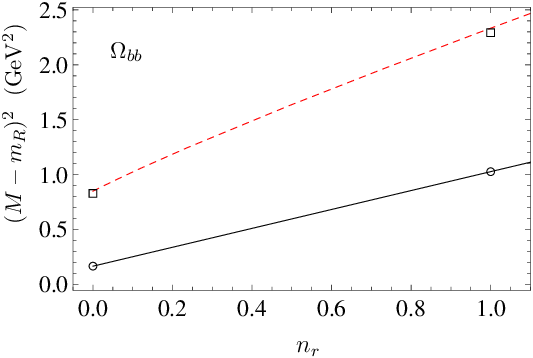}}
\subfigure[]{\label{subfigure:fiterr}\includegraphics[scale=0.45]{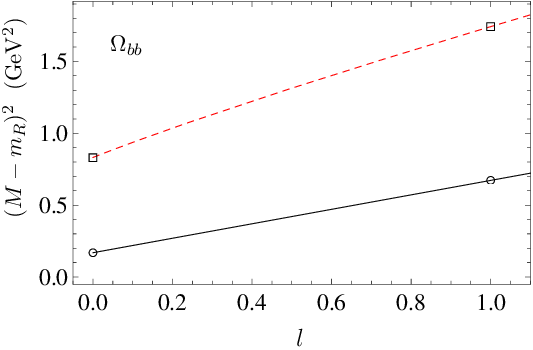}}
\caption{Same as Fig. \ref{fig:mr} except for the heavy-light baryons and Tables \ref{tab:bm} and \ref{tab:bmt}. }\label{fig:br}
\end{figure*}

In the diquark picture, the baryons consisting of one light quark and one heavy diquark or consisting of one heavy quark and one light diquark are denoted as the heavy-light baryons and belong to the heavy-light systems.
Employing (\ref{rtmeson}) with (\ref{rtft}) and (\ref{mrtf}) with (\ref{mrfp}) to the heavy-light baryons, the {\rts} are fitted. The used data are listed in Table \ref{tab:bm} and the fitted values are in Table \ref{tab:fitparameters}. The fitted {\rts} are in Fig. \ref{fig:br}.

\begin{figure*}[!phtb]
\centering
\subfigure[]{\label{subfigure:fiterr}\includegraphics[scale=0.47]{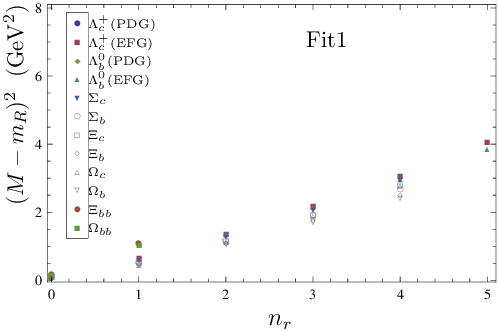}}
\subfigure[]{\label{subfigure:fiterr}\includegraphics[scale=0.47]{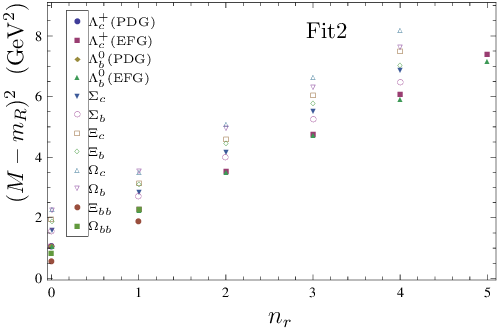}}
\subfigure[]{\label{subfigure:fiterr}\includegraphics[scale=0.47]{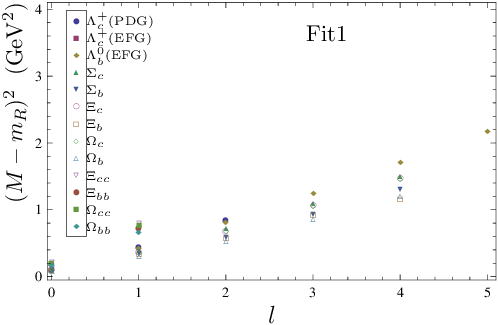}}
\subfigure[]{\label{subfigure:fiterr}\includegraphics[scale=0.47]{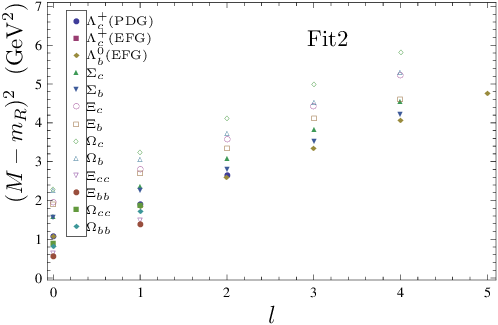}}
\caption{Same as Fig. \ref{fig:mc} except for the heavy-light baryons and Tables \ref{tab:bm} and \ref{tab:bmt}. }\label{fig:bc}
\end{figure*}

The difference of $m_R$ for two {\rt} relation leads to that the {\rts} (the red dashed lines) obtained by employing (\ref{mrtf}) with (\ref{mrfp}) lie above the {\rts} (the black lines) obtained by applying (\ref{rtmeson}) with (\ref{rtft}), see Fig. \ref{fig:br}.
From Fig. \ref{fig:bc} and the fitted values in Table \ref{tab:fitparameters}, we can see that the heavy-light baryons satisfy both of these two {\rt} relations.
We also notice that some fitted {\rts} by employing (\ref{rtmeson}) with (\ref{rtft}) are better than the fitted {\rts} by applying (\ref{rtmeson}) with (\ref{rtft}), for example, the {\rt} for $\Omega_c$, see \ref{subfigure:bocr}.

Similar to the meson case, from Fig. \ref{fig:bc} and the fitted values in Table \ref{tab:fitparameters}, we can see that both of the radial and orbital {\rts} for the heavy-light baryons plotted in the $((M-m_1-m_2-C)^2,\,x)$ planes almost overlap with each other irrespective of both the light constituent and the heavy constituent. This shows an universal description of these heavy-light baryons.
For the heavy-light baryons, the {\rts} in the $((M-m_1-C)^2,\,x)$ plane are universal only irrespective of heavy constituent.

As employing Eq. (8) with (12), the average values of $c_{fn_r}$ and $c_{fl}$ are $0.794$ and $0.749$ for the heavy-light baryons, respectively. When applying Eq. (14) with (15), the average values of $c_{fn_r}$ and $c_{fl}$ are $0.896$ and $0.855$, respectively.
The fitted $c_{fn_r}$ and $c_{fl}$ for the heavy-light baryons are smaller than that for the heavy-light mesons.
The fitted values for the doubly heavy baryons approximate the fitted values for the heavy-light mesons because the doubly heavy baryons strongly resemble the heavy-light mesons.

\begin{table*}[!phtb]
\caption{The theoretical values (EFG) \cite{Ebert:2011kk,Ebert:2002ig} for the heavy-light baryons. The values are in {\gev}. }  \label{tab:bmt}
\centering
\begin{tabular*}{0.9\textwidth}{@{\extracolsep{\fill}}ccccccccccc@{}}
\hline\hline
        & $\Sigma_c$ &$\Sigma_b$ & $\Xi_c$  & $\Xi_b$  & $\Omega_c$  & $\Omega_b$ & $\Xi_{cc}$ & $\Xi_{bb}$ & $\Omega_{cc}$ &$\Omega_{bb}$ \\
\hline
$1S(\frac{3}{2})^+$    &2.519  &5.834 &2.649   &5.963   &2.768  &6.088 &3.727  &10.237  &3.872  &10.389      \\
$2S(\frac{3}{2})^+$   &2.936   &6.226 &3.026   &6.342   &3.123  &6.461 &       &10.860  &       &10.992\\
$3S(\frac{3}{2})^+$   &3.293   &6.583 &3.396   &6.695   &3.510  &6.811\\
$4S(\frac{3}{2})^+$   &3.598   &6.876 &3.709   &6.984   &3.830  &7.096\\
$5S(\frac{3}{2})^+$   &3.873   &7.129 &3.989   &7.234   &4.114  &7.343\\
$1P(\frac{5}{2})^-$    &2.789  &6.084 &2.929   &6.226   &3.051  &6.334 &4.155  &10.661  &4.303  &10.798      \\
$1D(\frac{7}{2})^+$    &3.013  &6.260 &3.147   &6.414   &3.283  &6.517 &  &  &  &      \\
$1F(\frac{9}{2})^-$    &3.209  &6.459 &3.357   &6.610   &3.485  &6.713 &  &  &  &      \\
$1G(\frac{11}{2})^+$   &3.386  &6.635 &3.536   &6.728   &3.665  &6.884 &  &  &  &      \\
\hline
\end{tabular*}
\end{table*}

\subsection{Heavy-light tetraquarks}

The heavy-light tetraquarks denote tetraquarks consisting of one heavy diquark and one light antidiquark or consisting of one light diquark and one heavy antidiquark. The tetraquarks composed of a diquark and antidiquark in color $\bar{3}$ and $3$ configurations are considered.

\begin{figure*}[!phtb]
\centering
\subfigure[]{\label{subfigure:fiterr}\includegraphics[scale=0.45]{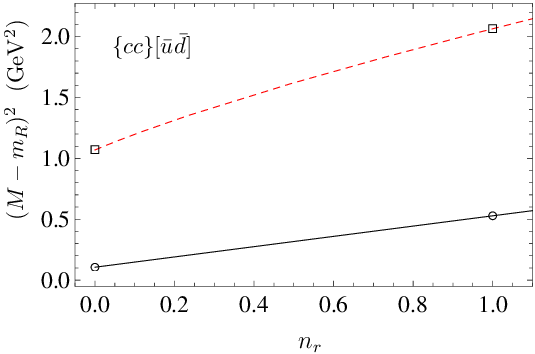}}
\subfigure[]{\label{subfigure:fiterr}\includegraphics[scale=0.45]{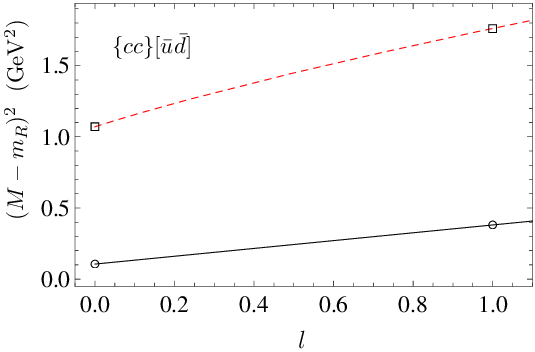}}
\subfigure[]{\label{subfigure:fiterr}\includegraphics[scale=0.45]{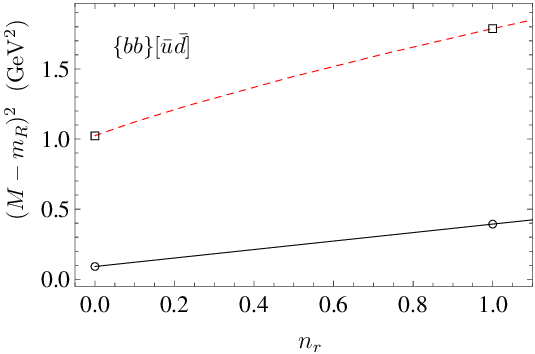}}
\subfigure[]{\label{subfigure:fiterr}\includegraphics[scale=0.45]{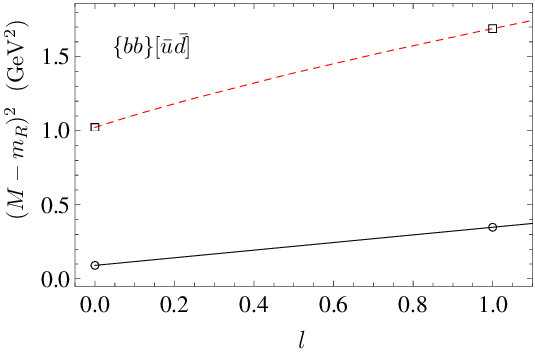}}
\subfigure[]{\label{subfigure:fiterr}\includegraphics[scale=0.45]{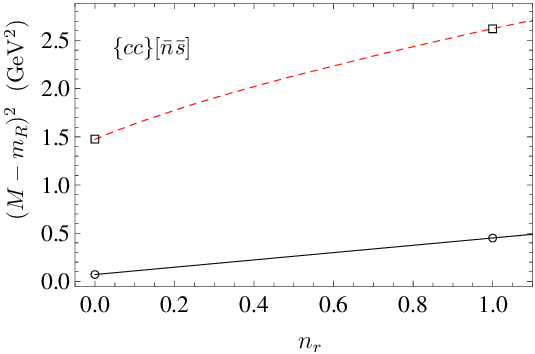}}
\subfigure[]{\label{subfigure:fiterr}\includegraphics[scale=0.45]{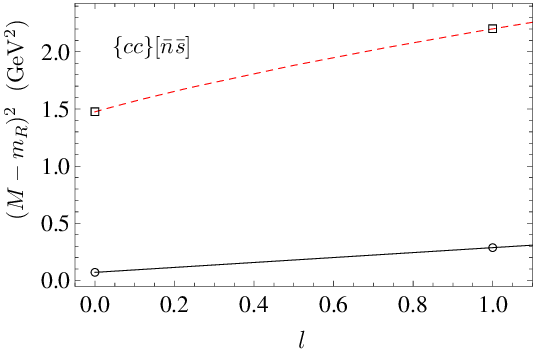}}
\subfigure[]{\label{subfigure:fiterr}\includegraphics[scale=0.45]{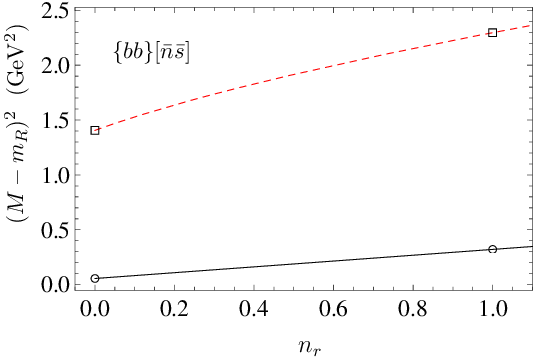}}
\subfigure[]{\label{subfigure:fiterr}\includegraphics[scale=0.45]{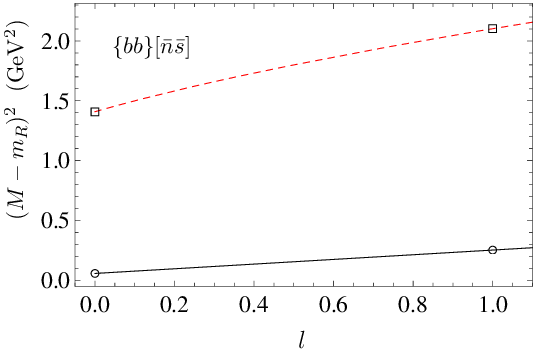}}
\subfigure[]{\label{subfigure:fiterr}\includegraphics[scale=0.45]{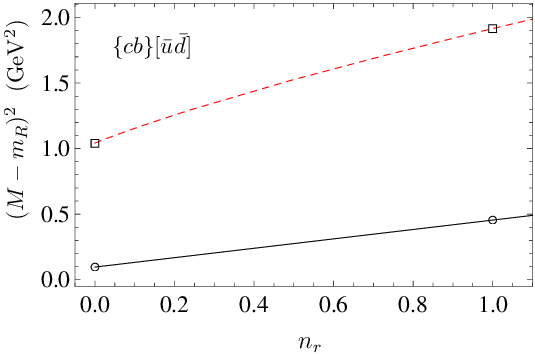}}
\subfigure[]{\label{subfigure:fiterr}\includegraphics[scale=0.45]{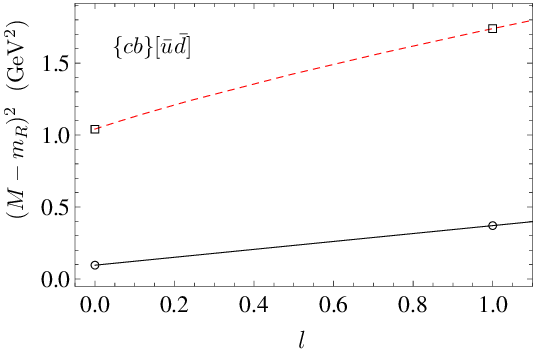}}
\subfigure[]{\label{subfigure:fiterr}\includegraphics[scale=0.45]{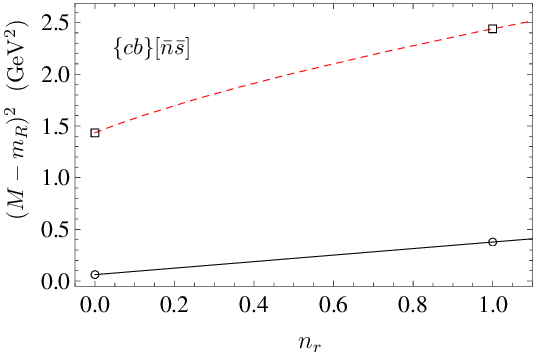}}
\subfigure[]{\label{subfigure:fiterr}\includegraphics[scale=0.45]{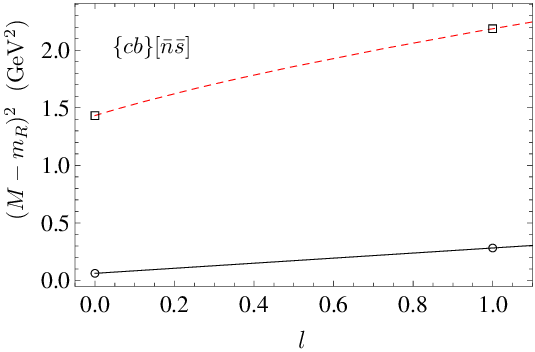}}
\subfigure[]{\label{subfigure:fiterr}\includegraphics[scale=0.45]{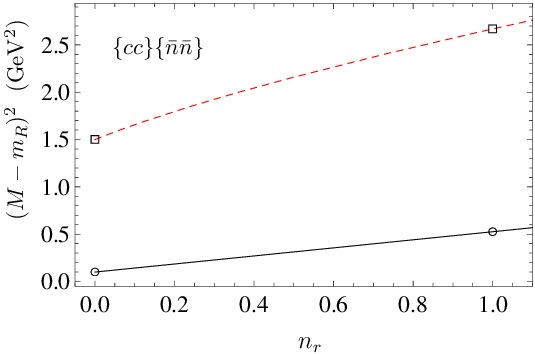}}
\subfigure[]{\label{subfigure:fiterr}\includegraphics[scale=0.45]{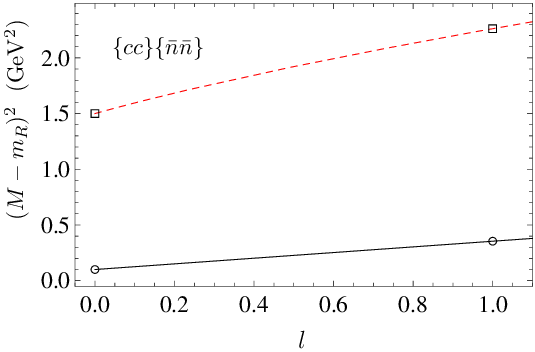}}
\subfigure[]{\label{subfigure:fiterr}\includegraphics[scale=0.45]{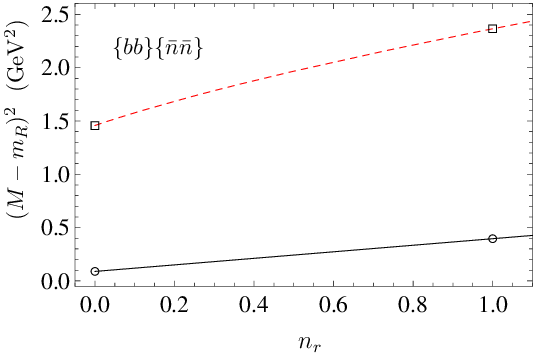}}
\subfigure[]{\label{subfigure:fiterr}\includegraphics[scale=0.45]{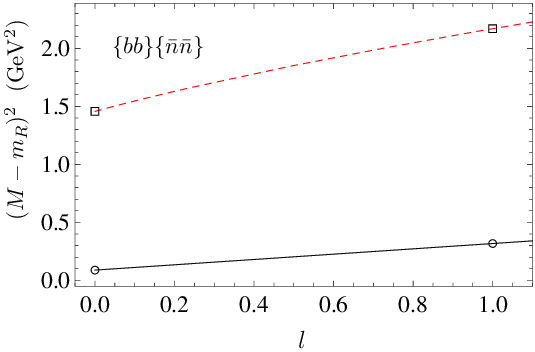}}
\subfigure[]{\label{subfigure:fiterr}\includegraphics[scale=0.45]{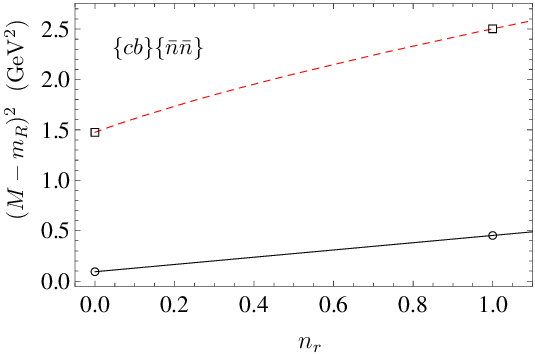}}
\subfigure[]{\label{subfigure:fiterr}\includegraphics[scale=0.45]{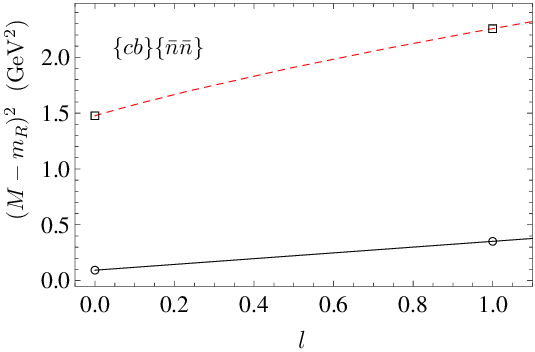}}
\caption{Same as Fig. \ref{fig:mr} except for the heavy-light tetraquarks and Table \ref{tab:tm}.}\label{fig:tr}
\end{figure*}

Applying Eqs. (\ref{rtmeson}) with (\ref{rtft}) and (\ref{mrtf}) with (\ref{mrfp}) and using data in Table \ref{tab:tm}, the radial and orbital {\rts} for the heavy-light tetraquarks are obtained, see Fig. \ref{fig:tr}. The fitted values are listed in Table \ref{tab:fitparameters}.
Similar to the heavy-light mesons and the heavy-light baryons, the {\rts} (the red dashed lines) obtained by employing (\ref{mrtf}) with (\ref{mrfp}) lie above the {\rts} (the black lines) obtained by applying (\ref{rtmeson}) with (\ref{rtft}), see Fig. \ref{fig:tr}.

\begin{figure*}[!phtb]
\centering
\subfigure[]{\label{subfigure:fiterr}\includegraphics[scale=0.47]{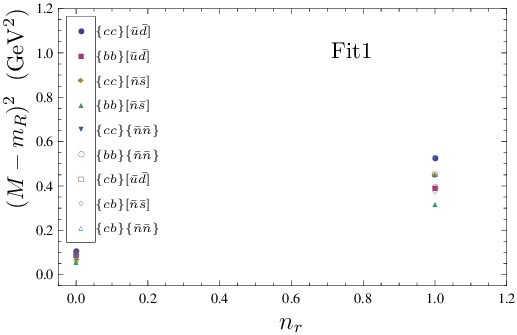}}
\subfigure[]{\label{subfigure:fiterr}\includegraphics[scale=0.47]{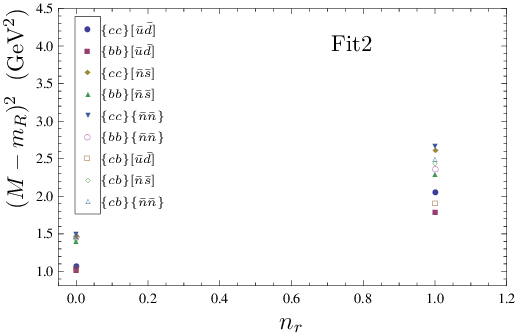}}
\subfigure[]{\label{subfigure:fiterr}\includegraphics[scale=0.47]{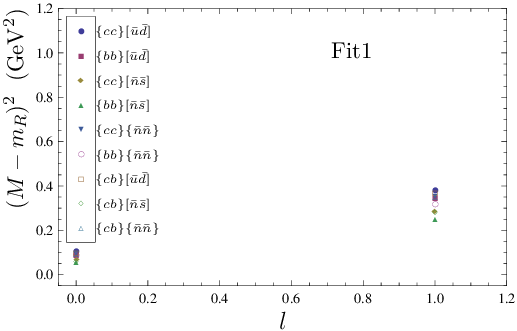}}
\subfigure[]{\label{subfigure:fiterr}\includegraphics[scale=0.47]{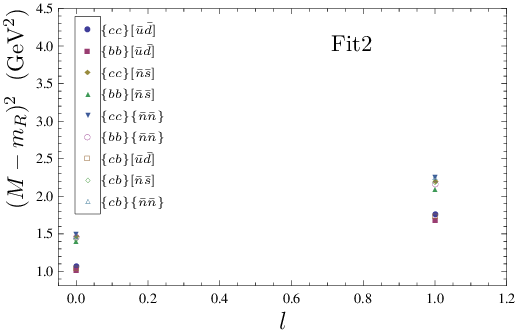}}
\caption{Same as Fig. \ref{fig:mc} except for the heavy-light tetraquarks and Table \ref{tab:tm}. }\label{fig:tc}
\end{figure*}

From Fig. \ref{fig:tc} and the fitted values in Table \ref{tab:fitparameters},
we can see that the heavy-light tetraquarks satisfy both of these two {\rt} relations.
Moreover, we can see that the universal description of the heavy-light tetraquarks approximately holds. We notice that the fitted results are not reliable enough because the experimental data is lack and the theoretical data on the $\lambda$-excited states are scarce. [The theoretical data on the $\lambda$-excited states of the tetraquarks are scarce because the masses of these states lie above the corresponding thresholds.]

As employing Eq. (\ref{rtmeson}) with (\ref{rtft}), the average values of $c_{fn_r}$ and $c_{fl}$ are $0.553$ and $0.579$ for the heavy-light tetraquarks, respectively. When applying Eq. (\ref{mrtf}) with (\ref{mrfp}), the mean values of $c_{fn_r}$ and $c_{fl}$ are $0.647$ and  $0.676$, respectively.
The fitted $c_{fn_r}$ and $c_{fl}$ for the heavy-light tetraquarks are smaller than that for the heavy-light mesons and the heavy-light baryons.

\begin{table*}[!phtb]
\caption{The theoretical values (KOS) \cite{Kim:2022mpa} for the doubly heavy tetraquarks. The values are in {\gev}. }  \label{tab:tm}
\centering
\begin{tabular*}{0.9\textwidth}{@{\extracolsep{\fill}}cccccccccc@{}}
\hline\hline
        & $\{cc\}[\bar{u}\bar{d}]$ &$\{bb\}[\bar{u}\bar{d}]$ & $\{cc\}[\bar{n}\bar{s}]$ &$\{bb\}[\bar{n}\bar{s}]$ & $\{cc\}\{\bar{n}\bar{n}\}$ & $\{bb\}\{\bar{n}\bar{n}\}$ & $\{cb\}[\bar{u}\bar{d}]$ & $\{cb\}[\bar{n}\bar{s}]$ &$\{cb\}\{\bar{n}\bar{n}\}$ \\
\hline
$1S$    &3.961  &10.489 &4.141   &10.664   &4.151  &10.685 &7.246  &7.423  &7.441       \\
$2S$    &4.363  &10.815 &4.545   &10.993  &4.560  &11.016 &7.610  &7.788  &7.808       \\
$1P$    &4.253  &10.778 &4.410   &10.928  &4.430  &10.951  &7.545 &7.705   &7.728    \\
\hline
\end{tabular*}
\end{table*}

\section{Concavity of the {\rts}}\label{sec:partners}

The {\rts} take different forms in different energy regions \cite{Chen:2021kfw,Chen:2022flh}. In this section, the {\rts} are depicted in the $(M^2,\,x)$ $(x=n_r,\,l)$ planes.
In Refs. \cite{Feng:2023txx,Chen:2023cws,Chen:2023ngj}, it is shown that the {\rts} for all types of diquarks (including the doubly heavy diquarks, the heavy-light diquarks and the light diquarks) are concave downwards.

In Ref. \cite{Chen:2018bbr}, it is shown that the meson {\rts} are concave downwards for the doubly heavy mesons and the heavy-light mesons. In the case of the light mesons, the {\rts} assume a linear form when the masses of the light quark and antiquark are taken as zero. However, upon taking into account the masses of the light quark and antiquark, the {\rts} for the light mesons also adopt a concave shape, see Refs. \cite{Selem:2006nd,Sonnenschein:2018fph}.

In Ref. \cite{Chen:2023djq}, we show that the {\rts} for the heavy-heavy baryons and tetraquarks are also concave downwards. In this work, we observe that the {\rts} for the heavy-light baryons and the heavy-light tetraquark also display a concave shape.
Taking into account the masses of the light constituent, it is expected that the {\rts} for the light baryons and the light tetraquarks will similarly be concave.

The curvature of the {\rts} holds significant importance \cite{Chen:2018bbr}. In potential models, the curvature is related to the dynamic equation and the confining potential.
Based on the discussions in this section, we assume that, when considering the mass of the light constituent, all {\rts} for the diquarks, mesons, baryons and tetraquarks are concave downwards in the $(M^2,\,n_r)$ and $(M^2,\,l)$ planes. Future experimental data will either validate or challenge the concavity conjecture. It will improve understanding of hadron dynamics and promote the study of hadron spectra.

\section{Conclusions}\label{sec:concl}
We apply two newly proposed {\rt} relations (\ref{rtmeson}) along with (\ref{rtft}) and (\ref{mrtf}) along with (\ref{mrfp}) to analyze the heavy-light systems, presenting the numerical plots for clarity. We find that the heavy-light diquarks, the heavy-light mesons, the heavy-light baryons and the heavy-light tetraquarks satisfy these two formulas.

In the scenario where Eq. (\ref{rtmeson}) is combined with (\ref{rtft}), the heavy-light systems conform to the universal description regardless of the masses of the light constituents and the heavy constituent. When considering Eq. (\ref{mrtf}) alongside (\ref{mrfp}), the heavy-light systems adhere to the universal description regardless of the masses of the heavy constituents. It is observed that the fitted slopes for the radial Regge trajectories are greater than those for the orbital Regge trajectories across all heavy-light systems.

The fitted slopes exhibit distinctive differences among the heavy-light mesons, baryons and tetraquarks, respectively. When employing Eq. (\ref{rtmeson}) with (\ref{rtft}), the average values of $c_{fn_r}$ ($c_{fl}$) are $1.026$, $0.794$ and $0.553$ ($1.026$, $0.749$ and $0.579$) for the heavy-light mesons, the heavy-light baryons and the heavy-light tetraquarks, respectively. When applying Eq. (\ref{mrtf}) with (\ref{mrfp}), the average values of $c_{fn_r}$ ($c_{fl}$) are $1.108$, $0.896$ and $0.647$ ($1.114$, $0.855$ and $0.676$) for the heavy-light mesons, baryons and tetraquarks, respectively.

Furthermore, the fitted results indicate that the {\rts} for the heavy-light systems exhibit concave downward shapes in the $(M^2,\,n_r)$ and $(M^2,\,l)$ planes. It is expected that, when considering the mass of the light constituent, all {\rts} for the diquarks, mesons, baryons and tetraquarks are concave downwards in the $(M^2,\,n_r)$ and $(M^2,\,l)$ planes. Future experimental data will either validate or refute the concavity conjecture. It will improve understanding of hadron dynamics and promote the study of hadron spectra.

\flushleft{\bf Acknowledgements}
We are very grateful to the anonymous referees for the valuable comments and suggestions.

\flushleft{\bf Data Availability Statement} This manuscript has no associated data or the data will not be deposited. [Authors' comment: All data are included in the manuscript.]

\flushleft
{\bf Code Availability Statement} The manuscript has no associated
code/software. [Author's comment: The manuscript has no associated
code.]

\flushleft
{\bf Open Access} This article is licensed under a Creative Commons Attribution
4.0 International License,which permits use, sharing, adaptation, distribution and reproduction in any medium or format, as long as you
give appropriate credit to the original author(s) and the source, provide
a link to the Creative Commons licence, and indicate if changes
were made. The images or other third party material in this article
are included in the article's Creative Commons licence, unless indicated
otherwise in a credit line to the material. If material is not
included in the article's Creative Commons licence and your intended
use is not permitted by statutory regulation or exceeds the permitted
use, you will need to obtain permission directly from the copyright
holder. To view a copy of this licence, visit \url{http://creativecommons.org/licenses/by/4.0/}.
Funded by SCOAP$^3$.


\begin{thebibliography}{99}

\bibitem{ParticleDataGroup:2022pth}
R.~L.~Workman \textit{et al.} [Particle Data Group],
PTEP \textbf{2022}, 083C01 (2022) and 2023 update
doi:10.1093/ptep/ptac097


\bibitem{Regge:1959mz}
  T.~Regge,
  Nuovo Cim.\  {\bf 14}, 951 (1959).


\bibitem{Chew:1961ev}
  G.~F.~Chew and S.~C.~Frautschi,
  Phys.\ Rev.\ Lett.\  {\bf 7}, 394 (1961).


\bibitem{Chew:1962eu}
  G.~F.~Chew and S.~C.~Frautschi,
  Phys.\ Rev.\ Lett.\  {\bf 8}, 41 (1962).



\bibitem{Collins:1971ff}
  P.~D.~B.~Collins,
  Phys.\ Rept.\  {\bf 1}, 103 (1971).


\bibitem{Collins:1977jy}
  P.~D.~B.~Collins,
  \emph{An Introduction to Regge Theory and High-Energy Physics} (Cambridge University Press, London, 1977).


\bibitem{Gross:2022hyw}
F.~Gross, E.~Klempt, S.~J.~Brodsky, A.~J.~Buras, V.~D.~Burkert, G.~Heinrich, K.~Jakobs, C.~A.~Meyer, K.~Orginos and M.~Strickland, \textit{et al.}
Eur. Phys. J. C \textbf{83}, 1125 (2023)
doi:10.1140/epjc/s10052-023-11949-2
[arXiv:2212.11107 [hep-ph]].


\bibitem{Inopin:2001ub}
  A.~E.~Inopin,
  arXiv: hep-ph/0110160, and references therein.


\bibitem{Inopin:1999nf}
  A.~Inopin and G.~S.~Sharov,
  Phys.\ Rev.\ D {\bf 63}, 054023 (2001).
  arXiv: hep-ph/9905499.



\bibitem{Guo:2008he}
X.~H.~Guo, K.~W.~Wei and X.~H.~Wu,
Phys. Rev. D \textbf{78}, 056005 (2008)
doi:10.1103/PhysRevD.78.056005
[arXiv:0809.1702 [hep-ph]].

\bibitem{Feng:2022hpj}
X.~C.~Feng, K.~W.~Wei, Jie-Wu and J.~Wu,
Eur. Phys. J. A \textbf{58}, no.11, 233 (2022)
doi:10.1140/epja/s10050-022-00886-5
[arXiv:2211.07083 [hep-ph]].



\bibitem{MartinContreras:2023oqs}
M.~A.~Martin Contreras and A.~Vega,
Phys. Rev. D \textbf{108}, no.12, 126024 (2023)
doi:10.1103/PhysRevD.108.126024
[arXiv:2309.02905 [hep-ph]].


\bibitem{Brisudova:1999ut}
M.~M.~Brisudova, L.~Burakovsky and J.~T.~Goldman,
Phys. Rev. D \textbf{61}, 054013 (2000)
doi:10.1103/PhysRevD.61.054013
[arXiv:hep-ph/9906293 [hep-ph]].


\bibitem{Sergeenko:1994ck}
M.~N.~Sergeenko,
Z. Phys. C \textbf{64}, 315-322 (1994)
doi:10.1007/BF01557404


\bibitem{Veseli:1996gy}
S.~Veseli and M.~G.~Olsson,
Phys. Lett. B \textbf{383}, 109-115 (1996)
doi:10.1016/0370-2693(96)00721-6
[arXiv:hep-ph/9606257 [hep-ph]].



\bibitem{Afonin:2014nya}
S.~S.~Afonin and I.~V.~Pusenkov,
Phys. Rev. D \textbf{90}, no.9, 094020 (2014)
doi:10.1103/PhysRevD.90.094020
[arXiv:1411.2390 [hep-ph]].


\bibitem{Burns:2010qq}
T.~J.~Burns, F.~Piccinini, A.~D.~Polosa and C.~Sabelli,
Phys. Rev. D \textbf{82}, 074003 (2010)
doi:10.1103/PhysRevD.82.074003
[arXiv:1008.0018 [hep-ph]].


\bibitem{Tang:2000tb}
A.~Tang and J.~W.~Norbury,
Phys. Rev. D \textbf{62}, 016006 (2000)
doi:10.1103/PhysRevD.62.016006
[arXiv:hep-ph/0004078 [hep-ph]].

\bibitem{Baldicchi:1998gt}
M.~Baldicchi and G.~M.~Prosperi,
Phys. Lett. B \textbf{436}, 145-152 (1998)
doi:10.1016/S0370-2693(98)00830-2
[arXiv:hep-ph/9803390 [hep-ph]].


\bibitem{Badalian:2019lyz}
A.~M.~Badalian and B.~L.~G.~Bakker,
Phys. Rev. D \textbf{100}, no.3, 034010 (2019)
doi:10.1103/PhysRevD.100.034010
[arXiv:1901.10280 [hep-ph]].


\bibitem{Chen:2018nnr}
J.~K.~Chen,
Eur. Phys. J. C \textbf{78}, no.8, 648 (2018)
doi:10.1140/epjc/s10052-018-6134-0


\bibitem{MartinContreras:2020cyg}
M.~A.~Martin Contreras and A.~Vega,
Phys. Rev. D \textbf{102} (2020) no.4, 046007
doi:10.1103/PhysRevD.102.046007
[arXiv:2004.10286 [hep-ph]].




\bibitem{Chen:2016spr}
  H.~X.~Chen, W.~Chen, X.~Liu, Y.~R.~Liu and S.~L.~Zhu,
  Rept.\ Prog.\ Phys.\  {\bf 80}, no. 7, 076201 (2017).
  arXiv: hep-ph/1609.08928.


\bibitem{Klempt:2012fy}
E.~Klempt and B.~C.~Metsch,
Eur. Phys. J. A \textbf{48}, 127 (2012)
doi:10.1140/epja/i2012-12127-1


\bibitem{Brodsky:2018vyy}
S.~J.~Brodsky,
Few Body Syst. \textbf{59}, no.5, 83 (2018)
doi:10.1007/s00601-018-1409-4
[arXiv:1802.08552 [hep-ph]].


\bibitem{Nielsen:2018uyn}
M.~Nielsen and S.~J.~Brodsky,
Phys. Rev. D \textbf{97}, no.11, 114001 (2018)
doi:10.1103/PhysRevD.97.114001
[arXiv:1802.09652 [hep-ph]].


\bibitem{Sonnenschein:2018fph}
J.~Sonnenschein and D.~Weissman,
Eur. Phys. J. C \textbf{79}, no.4, 326 (2019)
doi:10.1140/epjc/s10052-019-6828-y
[arXiv:1812.01619 [hep-ph]].



\bibitem{Chen:2017fcs}
K.~Chen, Y.~Dong, X.~Liu, Q.~F.~L\"u and T.~Matsuki,
Eur. Phys. J. C \textbf{78}, no.1, 20 (2018)
doi:10.1140/epjc/s10052-017-5512-3
[arXiv:1709.07196 [hep-ph]].



\bibitem{Sharov:2013tga}
G.~S.~Sharov,
[arXiv:1305.3985 [hep-ph]].


\bibitem{Forkel:2007cm}
H.~Forkel, M.~Beyer and T.~Frederico,
JHEP \textbf{07}, 077 (2007)
doi:10.1088/1126-6708/2007/07/077
[arXiv:0705.1857 [hep-ph]].

\bibitem{A:2023bxv}
C.~M.~A. and R.~Dhir,
[arXiv:2311.05274 [hep-ph]].


\bibitem{Chen:2021kfw}
J.~K.~Chen,
Eur. Phys. J. A \textbf{57}, 238 (2021)
doi:10.1140/epja/s10050-021-00502-y
[arXiv:2102.07993 [hep-ph]].


\bibitem{Chen:2022flh}
J.~K.~Chen,
Nucl. Phys. B \textbf{983}, 115911 (2022)
doi:10.1016/j.nuclphysb.2022.115911
[arXiv:2203.02981 [hep-ph]].



\bibitem{Chen:2023djq}
J.~K.~Chen,
[arXiv:2302.05926 [hep-ph]].


\bibitem{Feng:2023txx}
X.~Feng, J.~K.~Chen and J.~Q.~Xie,
Phys. Rev. D \textbf{108}, no.3, 034022 (2023)
doi:10.1103/PhysRevD.108.034022
[arXiv:2305.15705 [hep-ph]].


\bibitem{Chen:2023cws}
J.~K.~Chen, X.~Feng and J.~Q.~Xie,
JHEP \textbf{10}, 052 (2023)
doi:10.1007/JHEP10(2023)052
[arXiv:2308.02289 [hep-ph]].




\bibitem{Ferretti:2019zyh}
J.~Ferretti,
Few Body Syst. \textbf{60}, no.1, 17 (2019)
doi:10.1007/s00601-019-1483-2

\bibitem{Bedolla:2019zwg}
M.~A.~Bedolla, J.~Ferretti, C.~D.~Roberts and E.~Santopinto,
Eur. Phys. J. C \textbf{80}, no.11, 1004 (2020)
doi:10.1140/epjc/s10052-020-08579-3
[arXiv:1911.00960 [hep-ph]].


\bibitem{Godfrey:1985xj}
S.~Godfrey and N.~Isgur,
Phys. Rev. D \textbf{32}, 189-231 (1985)
doi:10.1103/PhysRevD.32.189

\bibitem{Durand:1981my}
B.~Durand and L.~Durand,
Phys. Rev. D \textbf{25}, 2312 (1982)
doi:10.1103/PhysRevD.25.2312

\bibitem{Durand:1983bg}
B.~Durand and L.~Durand,
Phys. Rev. D \textbf{30}, 1904 (1984)
doi:10.1103/PhysRevD.30.1904

\bibitem{Lichtenberg:1982jp}
D.~B.~Lichtenberg, W.~Namgung, E.~Predazzi and J.~G.~Wills,
Phys. Rev. Lett. \textbf{48}, 1653 (1982)
doi:10.1103/PhysRevLett.48.1653


\bibitem{Jacobs:1986gv}
S.~Jacobs, M.~G.~Olsson and C.~Suchyta, III,
Phys. Rev. D \textbf{33}, 3338 (1986)
[erratum: Phys. Rev. D \textbf{34}, 3536 (1986)]
doi:10.1103/PhysRevD.33.3338



\bibitem{Ferretti:2011zz}
J.~Ferretti, A.~Vassallo and E.~Santopinto,
Phys. Rev. C \textbf{83}, 065204 (2011)
doi:10.1103/PhysRevC.83.065204

\bibitem{Eichten:1974af}
E.~Eichten, K.~Gottfried, T.~Kinoshita, J.~B.~Kogut, K.~D.~Lane and T.~M.~Yan,
Phys. Rev. Lett. \textbf{34}, 369-372 (1975)
[erratum: Phys. Rev. Lett. \textbf{36}, 1276 (1976)]
doi:10.1103/PhysRevLett.34.369

\bibitem{Lucha:1991vn}
W.~Lucha, F.~F.~Schoberl and D.~Gromes,
Phys. Rept. \textbf{200}, 127-240 (1991)
doi:10.1016/0370-1573(91)90001-3


\bibitem{Gromes:1981cb}
D.~Gromes,
Z. Phys. C \textbf{11}, 147 (1981)
doi:10.1007/BF01573997


\bibitem{Selem:2006nd}
A.~Selem and F.~Wilczek,
doi:10.1142/9789812773524\_0030
[arXiv:hep-ph/0602128 [hep-ph]].

\bibitem{Chen:2014nyo}
B.~Chen, K.~W.~Wei and A.~Zhang,
Eur. Phys. J. A \textbf{51}, 82 (2015)
doi:10.1140/epja/i2015-15082-3
[arXiv:1406.6561 [hep-ph]].


\bibitem{Jakhad:2023mni}
P.~Jakhad, J.~Oudichhya, K.~Gandhi and A.~K.~Rai,
Phys. Rev. D \textbf{108}, no.1, 014011 (2023)
doi:10.1103/PhysRevD.108.014011
[arXiv:2306.06349 [hep-ph]].




\bibitem{brau:04bs}
 F.~Brau,
  Phys.\ Rev.\ D {\bf 62}, 014005 (2000).
  arXiv:hep-ph/0412170


\bibitem{brsom}
S. Tomonaga, \emph{Quantum Mechanics, Volume I: Old Quantum Theory}
(North-Holland Publishing Company, Amsterdam, 1962)



\bibitem{Jia:2018vwl}
D.~Jia and W.~C.~Dong,
Eur. Phys. J. Plus \textbf{134}, no.3, 123 (2019)
doi:10.1140/epjp/i2019-12474-8
[arXiv:1811.04214 [hep-ph]].

\bibitem{Ebert:2007rn}
D.~Ebert, R.~N.~Faustov, V.~O.~Galkin and W.~Lucha,
Phys. Rev. D \textbf{76}, 114015 (2007)
doi:10.1103/PhysRevD.76.114015
[arXiv:0706.3853 [hep-ph]].


\bibitem{Ebert:2011kk}
D.~Ebert, R.~N.~Faustov and V.~O.~Galkin,
Phys. Rev. D \textbf{84}, 014025 (2011)
doi:10.1103/PhysRevD.84.014025
[arXiv:1105.0583 [hep-ph]].


\bibitem{Ebert:2009ua}
D.~Ebert, R.~N.~Faustov and V.~O.~Galkin,
Eur. Phys. J. C \textbf{66}, 197-206 (2010)
doi:10.1140/epjc/s10052-010-1233-6
[arXiv:0910.5612 [hep-ph]].


\bibitem{Dosch:1988hu}
H.~G.~Dosch, M.~Jamin and B.~Stech,
Z. Phys. C \textbf{42}, 167 (1989)
doi:10.1007/BF01565139



\bibitem{Zhang:2006xp}
A.~Zhang, T.~Huang and T.~G.~Steele,
Phys. Rev. D \textbf{76}, 036004 (2007)
doi:10.1103/PhysRevD.76.036004
[arXiv:hep-ph/0612146 [hep-ph]].


\bibitem{Jamin:1989hh}
M.~Jamin and M.~Neubert,
Phys. Lett. B \textbf{238}, 387-394 (1990)
doi:10.1016/0370-2693(90)91753-X


\bibitem{deOliveira:2023hma}
T.~de Oliveira, D.~Harnett, R.~Kleiv, A.~Palameta and T.~G.~Steele,
Phys. Rev. D \textbf{108}, no.5, 054036 (2023)
doi:10.1103/PhysRevD.108.054036
[arXiv:2307.15815 [hep-ph]].


\bibitem{Kleiv:2013dta}
R.~T.~Kleiv, T.~G.~Steele, A.~Zhang and I.~Blokland,
Phys. Rev. D \textbf{87}, no.12, 125018 (2013)
doi:10.1103/PhysRevD.87.125018
[arXiv:1304.7816 [hep-ph]].


\bibitem{Wang:2011ab}
Z.~G.~Wang,
Commun. Theor. Phys. \textbf{59}, 451-456 (2013)
doi:10.1088/0253-6102/59/4/11
[arXiv:1112.5910 [hep-ph]].


\bibitem{Esau:2019hqw}
S.~Esau, A.~Palameta, R.~T.~Kleiv, D.~Harnett and T.~G.~Steele,
Phys. Rev. D \textbf{100}, 074025 (2019)
doi:10.1103/PhysRevD.100.074025
[arXiv:1905.12803 [hep-ph]].


\bibitem{Wang:2010sh}
Z.~G.~Wang,
Eur. Phys. J. C \textbf{71}, 1524 (2011)
doi:10.1140/epjc/s10052-010-1524-y
[arXiv:1008.4449 [hep-ph]].


\bibitem{Ebert:2002ig}
D.~Ebert, R.~N.~Faustov, V.~O.~Galkin and A.~P.~Martynenko,
Phys. Rev. D \textbf{66}, 014008 (2002)
doi:10.1103/PhysRevD.66.014008
[arXiv:hep-ph/0201217 [hep-ph]].

\bibitem{Kim:2022mpa}
Y.~Kim, M.~Oka and K.~Suzuki,
Phys. Rev. D \textbf{105}, no.7, 074021 (2022)
doi:10.1103/PhysRevD.105.074021
[arXiv:2202.06520 [hep-ph]].


\bibitem{Chen:2023ngj}
J.~K.~Chen, J.~Q.~Xie, X.~Feng and H.~Song,
Eur. Phys. J. C \textbf{83}, no.12, 1133 (2023)
doi:10.1140/epjc/s10052-023-12329-6
[arXiv:2310.05131 [hep-ph]].


\bibitem{Chen:2018bbr}
J.~K.~Chen,
Phys. Lett. B \textbf{786}, 477-484 (2018)
doi:10.1016/j.physletb.2018.10.022
[arXiv:1807.11003 [hep-ph]].





\end{thebibliography}
\end{document}